\documentclass{article}
\usepackage{amsfonts,amssymb, amsmath}

\usepackage[cp1251]{inputenc}
\usepackage[T2A]{fontenc}
\usepackage[english]{babel}
\usepackage{graphicx}

\newtheorem{prop}{Proposition}
\makeatletter
\newcommand\xleftrightarrow[2][]{\ext@arrow 0099{\longleftrightarrowfill@}{#1}{#2}}
\def\longleftrightarrowfill@{\arrowfill@\leftarrow\relbar\rightarrow}
\makeatother

\def\nn{\nonumber }
\def\bq{ \begin{equation}}
\def\eq{ \end{equation}}
\def\ben{ \begin{eqnarray}}
\def\en{ \end{eqnarray}}
\def\a{{\alpha}}
\def\b{{\beta}}

\begin{document}


\title{On discretization of the Euler top}
\author{A.V. Tsiganov \\
\it\small St.Petersburg State University, St.Petersburg, Russia;  Udmurt State University,  Izhevsk,  Russia\\
\it\small e--mail:  andrey.tsiganov@gmail.com}
\date{}
\maketitle
\begin{abstract}
Application of the intersection theory to construction of $n$-point finite-difference equations associated with  classical integrable systems is discussed.  As an example, we present a few new discretizations of  motion of the Euler top sharing the integrals of motion with the continuous time system and the  Poisson bracket up to the integer scaling  factor.
 \end{abstract}

\section{Introduction}
\setcounter{equation}{0}
This paper deals with discretization equations of motion for one of the basic integrable systems, the three-dimensional Euler top, which describes  motion of a free rigid body with a fixed point
\bq\label{eul-eqV}
A\dot{p}+(C-B)qr=0\,,\quad
B\dot{q}+(A-C)rp=0\,,\quad
C\dot{r}+(B-A)pq=0\,.
\eq
Here $\Omega=(p,q,r)$ is the vector of angular velocity in the
coordinate system attached firmly to the body; axes of this system coincide with the principal axes of inertia, and numbers $A,B,C>0$ are the corresponding principal  moments of inertia.

In the world coordinate system angular momentum  vector
\[M=(Ap,Bq,Cr)=(M_1,M_2,M_3)\]
 is fixed and its length in body coordinate system
 \begin{equation}\label{eul-k}
 K=(M,M)=A^2p^2+B^2q^2+C^2r^2\,,
 \end{equation}
is a first integral of equations (\ref{eul-eqV}). Another first integral is a kinetic energy for the body
 \begin{equation}\label{eul-h}
T=\frac12(\Omega,M)=\frac12\left(Ap^2+Bq^2+Cr^2\right)=\frac12\left(\dfrac{M_1^2}{A}+\dfrac{M_2^2}{B}+\dfrac{M_3^2}{C}\right)\;.
\end{equation}

The first discretization of the free $n$-dimensional top was constructed by Moser and Veselov \cite{mos91,ves88}
 by refactorization of matrix polynomials. This discretization is represented by an isospectral transformation
 \bq\label{eul-mv}
 \mathbf L_{k+1}=\mathbf A_k \mathbf L_k \mathbf A_k^{-1}
 \eq
 which does not explicitly involve a time step. Here $\mathbf L_k=\mathbf M_k+\lambda\mathbf  J^2$, $\mathbf A_k=\mathbf \Omega_k+\lambda \mathbf  J$, $\lambda$ is a spectral parameter and  $\mathbf  M, \mathbf \Omega,\mathbf  J$  are matrices associated with angular momentum, angular velocity and tensor of inertia, respectively.  Detailed exposition of this integrable discretization
 may be  found in the textbook \cite{sur03}.

In \cite{bob98} authors discuss the following  finite-difference equation
\[
\widetilde{M}_i-M_i =\gamma\alpha_i(\widetilde{M}_i + M_j)(\widetilde{M}_k +M_k)\,,
\]
where $\gamma$ is a function on phase space and $(ijk)$ stands for any cyclic permutation of (123). This implicit map $M\to\widetilde{M}$
 preserves first integrals of the continuous problem and  Poisson structure only for the special choice of function $\gamma$.

In \cite{hk00} explicit discrete map
\[
\widetilde{M}_i-M_i =\delta_i (\widetilde{M}_jM_k + M_j\widetilde{M}_k),
\]
where $\delta_i$ are some parameters, was constructed by applying the Hirota method. This map does not preserve first integrals of the equations (\ref{eul-eqV}), but has all the integrability attributes, i.e. there are two compatible invariant Poisson structures; two independent integrals of motion which are in involution with respect to any of the invariant Poisson brackets; a Lax representation; explicit solutions in terms of elliptic functions and so on, see \cite{sur10}.

In \cite{fed05}  discretization of the classical Euler top
\bq\label{eul-fed}
\widetilde{M}_i-M_i =\phi_i(M_1,M_2M_3)\,,
\eq
where $\phi _i$ are some special functions was obtained by using the Poinsot model of motion.  In this model   conditions $T=E$ (\ref{eul-h})  and  $K=k$ (\ref{eul-k}) define the inertia ellipsoid and it's confocal ellipsoid, respectively.  Path of the angular velocity vector in three dimensional space (polhode) is the intersection of these two ellipsoids, a closed quartic curve on the inertia ellipsoid. Point $P$ on a polhode corresponds to some partial solution $\Omega(t)$ of (\ref{eul-eqV}) at fixed time $t$. According to \cite{fed05} two points $P$ and $\widetilde{P}$ on the polhode are related by transformations (\ref{eul-fed}), which can be obtained using the Chasles theorem on  tangent lines to a geodesic on a quadric.

Instead of the Chasles theorem on contact curves, we propose to use the intersection theory.  Indeed, for a given  integrable system we can  consider  solutions of equations of motion at $t=t_1,t_2,\ldots$, which form a set of points $P_1,P_2,\ldots$ on a common level surface $X$ of integrals of motion $H_1,H_2,\ldots$.   Finite-difference  equation
\bq\label{fd-eq}
\mathcal Y\left(P_{k-\ell},\ldots,P_k,\ldots P_{k+m};k\right)=0\,,\qquad k\in\mathbb Z
\eq
or $\ell+m+1$-point map \cite{hiet16} can be considered as a correspondence between the intersection points of surface $X$ with an auxiliary curve $Y$. If $X$ is the algebraic surface, we can study standard configurations of  intersection points and the corresponding  finite-difference equations (\ref{fd-eq}) in the framework of the classical intersection theory.

 Because points $P_1,P_2,\ldots$ are solutions of equations of motion with fixed values of the first integrals  we may suppose that $\ell+m+1$-point maps (\ref{fd-eq}) preserve the integrals of motion defining surface $X$ and the Poisson brackets defining equations of motion. For the Euler top two point maps $\ell+m=1$  sharing the integrals of motion with the continuous time system and the  Poisson bracket are well-known, see  \cite{bob98,fed05,mos91,sur10,ves88} and references within. The main aim of this article is to discuss  construction of   finite-difference equations in the framework  of the classical  intersection theory and present a few finite-difference  equations for the Euler top at $\ell+m>1$ and two-point maps sharing original first integrals up to the integer scaling factor.

Intersection theory is at the heart of algebraic geometry and its history is rich and fascinating. To construct finite-difference equation for  the Euler case,  when generic level curves of $T$  and $K$ are elliptic curves,  it will be enough to use Abel's results and  Clebsch's geometric interpretation of these results based on algebraic theory of curves which can be found in the classical text books \cite{bak97,gr}.  Of course, this construction could also be  described using modern mathematical language of intersection theory  \cite{eh16,ful84,grif04,kl05}.

\section{The Euler top}
Equations (\ref{eul-eqV})  in terms of angular momenta look like
\begin{eqnarray}\label{eul-eq}
\dot{M}_1 & = & \left(\frac{1}{B}-\frac{1}{C}\right)M_2M_3=(\a_2-\a_3)M_2M_3\;,\nonumber\\
\dot{M}_2 & = & \left(\frac{1}{C}-\frac{1}{A}\right)M_3M_1=(\a_3-\a_1)M_1M_3\;,\\
\dot{M}_3 & = & \left(\frac{1}{A}-\frac{1}{B}\right)M_1M_2=(\a_1-\a_2)M_1M_2\;,\nonumber
\end{eqnarray}
where
\[\alpha_1=A^{-1}\,,\qquad \a_2=B^{-1}\,,\qquad \alpha_3=C^{-1}\,.\]
These equations are Hamiltonian with respect to   Poisson bracket
\begin{equation}\label{eul-pb}
\{M_1,M_2\}=M_3\;,\qquad \{M_2,M_3\}=M_1\;,\qquad \{M_3,M_1\}=M_2\;.
\end{equation}
and  Hamilton function  $H=2T$. The bracket (\ref{eul-pb}) is degenerate and its Casimir polynomial coincides with the first integral
$K=(M,M)$ (\ref{eul-k}).

 Let us express $M_{1,2}$ via third component of momenta $M_3$ and values of the  first integrals $2T=h$ and $K=k$:
 \bq\label{eul-M12}
 M_1 = \sqrt{\dfrac{(\a_3-\a_2)M_3^2+\a_2k-h)}{\a_2-\a_1}}\,,\qquad M_2 = \sqrt{\dfrac{(\a_1-\a_3)M_3^2-\a_1k+h)}{\a_2-\a_1}}\,.
 \eq
 Substituting these expressions into the third equation in (\ref{eul-eq})  and taking $M_3=x$ one gets standard  quadrature
\bq\label{eul-deq}
\int^{M_3}\frac{d {x}}{y}=2t\,,\qquad y^2=\Bigl((\a_1-\a_3)x^2-\a_1k+h\Bigr)\Bigl((\a_3-\a_2)x^2+\a_2k-h\Bigr)
\eq
on  elliptic curve $X$
\bq\label{ell-curve}
X:\quad y^2=f(x)\,,\qquad f(x)=a_4x^4+a_3x^3+a_2x^2+a_1x+a_0\,.
\eq
with coefficients
\bq\label{eul-coeff}
\begin{array}{rll}
a_4&=(\a_1-\a_3)(\a_3-\a_2)\,,\qquad &\a_3=0\,,\\
\\
 a_2&=(2\a_3-\a_1-\a_2)h+(2\a_1\a_2-\a_1\a_3-\a_2\a_3)k\,,\qquad &\a_1=0\,,\\
\\
 a_0&=(\a_1k-h)(h-\a_2k)\,.&
 \end{array}
\eq
Solution $M_3 (t) $ of the equation (\ref{eul-deq}) and, therefore, solutions of the original equations (\ref{eul-eq})  are expressed via Jacobi elliptic functions \cite{jac84}, see also \cite{gr}, p. 101-103. This solution  $M_3(t)$ of (\ref{eul-deq})  at $t=t_i$ is   point  $P_i$ on curve $X$ with abscissa $x_i=M_3(t_i)$ and ordinate  $y_i=\dot{M}_3(t_i)$.

Let $X$ be a smooth nonsingular algebraic  curve  on a projective plane over field $\mathrm k$.  Prime divisors are  points on $X$  denoted $P_i = (x_i, y_i)$,  and $P_\infty$, which is a point at infinity. Divisor
\[D = \sum m_iP_i\,,\qquad m_i\in \mathbb Z\]
is a formal sum of prime divisors, and the degree of  divisor $D$ is a sum deg$D=\sum m_i$ of  multiplicities of points in  support of the divisor.  Group of divisors Div$X$ is an additive Abelian group under the formal addition rule
 \[\sum m_i P_i+\sum n_i P_i=\sum (m_i+n_i) P_i\,.\]
 To define a linear equivalence relation on divisors we  can use the rational functions on $X$.
Function  $f$ is a quotient of two polynomials; they are each zero only on a finite closed subset of codimension one in $X$, which is therefore the union of finitely many prime divisors. The difference of these two subsets  define  a principal divisor $div f$
 associated with function $f$. The subgroup of Div$X$ consisting of the principal divisors is denoted by Prin$X$.
 So, two divisors $D, D'\in \mbox{Div} X$ are linearly equivalent
\[D\approx D'\]
if their difference $D-D'$ is principal divisor
\[
D-D'=div(g)\equiv 0\quad \mathrm{mod\, Prin}X\,,
\]
i.e. divisor of  rational function $g$ on $X$.    The Picard group of $X$ is  quotient group
\[
\mbox{Pic}X =\dfrac{\mbox{Div}X}{\mbox{Prin}X}=\dfrac{\mbox{Divisors defined over  k}}{\mbox{Divisors of functions defined over k}}\,.
\]
If $X$ is a “projective variety”, then the group Pic$X$ underlies a natural $\mathrm k$-scheme (Picard scheme), which is a disjoint union of quasi-projective schemes, and  the operations of multiplying and inverting are given by $\mathrm k$-maps  \cite{kl05}. Then we can extend the theory of schemes to stacks and so on \cite{eh16}.

 We can consider the finite-difference equation (\ref{fd-eq}), associated with quadrature (\ref{eul-deq}), as a correspondence between prime divisors on $X$
 \[
\mathcal Y\left(P_{k-\ell},\ldots, P_{k+m};k\right)=0\, \mathrm{mod\, Prin}X\,.
\]
 If we restrict ourselves by linear correspondences
\[
 \mathcal Y\left(P_{k-\ell},\ldots,P_k,\ldots, P_{k+m};k\right)=\sum_{i=k-\ell}^{k+m} m_iP_i= 0\, \mathrm{mod\, Prin}X\,,
\]
 these equations can be identified with an intersection divisor of $X$ with some auxiliary curve $Y$
\[
div(X\cdot Y)=0.
\]
It allows us to construct finite-difference equations  (\ref{fd-eq}) using well-known group operations,
operations on schemes and stacks \cite{eh16,ful84,kl05}.

\subsection{Examples of intersection divisors}
Let us consider intersection of  plane curve $X$ (\ref{ell-curve}) with a parabola
\[
Y:\qquad y=\mathcal P(x)\,,\qquad \mathcal P(x)=b_2x^2+b_1x+b_0
\]
and the corresponding intersection divisor $div(X\cdot Y)$ of degree four, see \cite{bak97}, p.113 or  \cite{gr}, p.166. Following Abel we substitute $y=\mathcal P(x)$ into (\ref{ell-curve}) and obtain the  so-called Abel polynomial
\[
\psi(x)=\mathcal P(x)^2-f(x)\,.
\]
Divisor of this polynomial on $X$ coincides with $div(X\cdot Y)$, i.e. roots of this polynomial are abscissas of  intersection points $P_1,P_2,P_3$ and $P_4$ forming support of the intersection divisor $div(X\cdot Y)$.

At $b_2=\sqrt{a_4}$ one of the intersection points is $P_\infty$. In this case $\psi(x)$ is equal to
\[\begin{array}{rcl}
\psi(x)&=&(2b_1b_2-a_3)x^3+(2b_0b_2+b_1^2-a_2)x^2+(2b_0b_1-a_1)x+b_0^2-a_0\nn\\
\nn\\
&=&(2b_1b_2-a_3)(x-x_1)(x-x_2)(x-x_3)\,.
\end{array}
\]
Equating coefficients of $\psi$  one gets relation between abscissas of  the remaining  points $P_1,P_2$ and $P_3$ in support of the intersection divisor
\bq\label{add-gen}
x_1+x_2+x_3=-\dfrac{2b_0b_2+b_1^2-a_2}{2b_1b_2-a_3}\,.
\eq
If $P_i\neq P_j$, we can define parabola $Y$  using the Lagrange interpolation by any pair of points  $(P_1,P_2)$, $(P_1,P_3)$ or $(P_2,P_3)$.
For instance, taking the following pair of points $(P_1,P_2)$ one gets
\[
\mathcal P(x)=b_2x^2+b_1x+b_0=\sqrt{a_4}(x-x_1)(x-x_2)+\dfrac{(x-x_2)y_1}{x_1-x_2}+\dfrac{(x-x_1)y_2}{x_2-x_1},
\]
which allows us to determine  $b_2,b_1,b_0$ as functions on $x_{1,2}$ and $y_{1,2}$. Substituting coefficients of $\mathcal P (x) $ into the equation (\ref{add-gen}) we obtain an explicit expression for abscissa  $x_3$  as a function on coordinates  $x_ {1,2} $ and $y_ {1,2} $
\bq\label{add-ell}
x_3=-x_1-x_2+\phi(x_1,y_1,x_2,y_2)\,,\qquad \phi=-\dfrac{2b_0b_2+b_1^2-a_2}{2b_1b_2-a_3}
\eq
Then we can calculate ordinate $y_3=\mathcal P (x_3) $  of the third intersection point $P_3$.

If we have a double intersection point, for instance $P_1=P_3$, then
\bq\label{doub-ell}
x_2=-2x_1+\phi(x_1,y_1)\,,\qquad \phi=-\dfrac{2b_0b_2+b_1^2-a_2}{2b_1b_2-a_3}\,,
\eq
where function $\phi(x_1,y_1)$ is defined by $\mathcal P(x)$ due to the  Hermite interpolation
\[
\mathcal P(x)=b_2x^2+b_1x+b_0=\sqrt{a_4}(x-x_1)^2+\dfrac{(x-x_1)(4a_4x_1^3+3a_3x_1^2+2a_2x_1+a_1)}{2y_1}+y_1\,.
\]

In modern terms, we consider two partitions of the intersection divisor
\[
div(X\cdot Y)=(P_1+P_2)+P_3+P_\infty\,,\qquad\mbox{and}\qquad
div(X\cdot Y)=(2P_1)+P_2+P_\infty.
\]
Here we use brackets $(.)$ in order to visually separate a part of  the intersection divisor which will be used for polynomial interpolation of auxiliary curve $Y$. Because
\[
div(X\cdot Y)=0\,,
\]
these partitions can be rewritten as addition and doubling of prime divisors
\[
-P_3=P_1+P_2\,,\qquad -P_2=2P_1\,,
\]
 where inversion is  $(x,y)\to (x,-y)$, see Figure 1.
\begin{figure}[!ht]
\begin{minipage}[h]{0.49\linewidth}
\center{\includegraphics[width=0.85\linewidth, height=0.2\textheight]{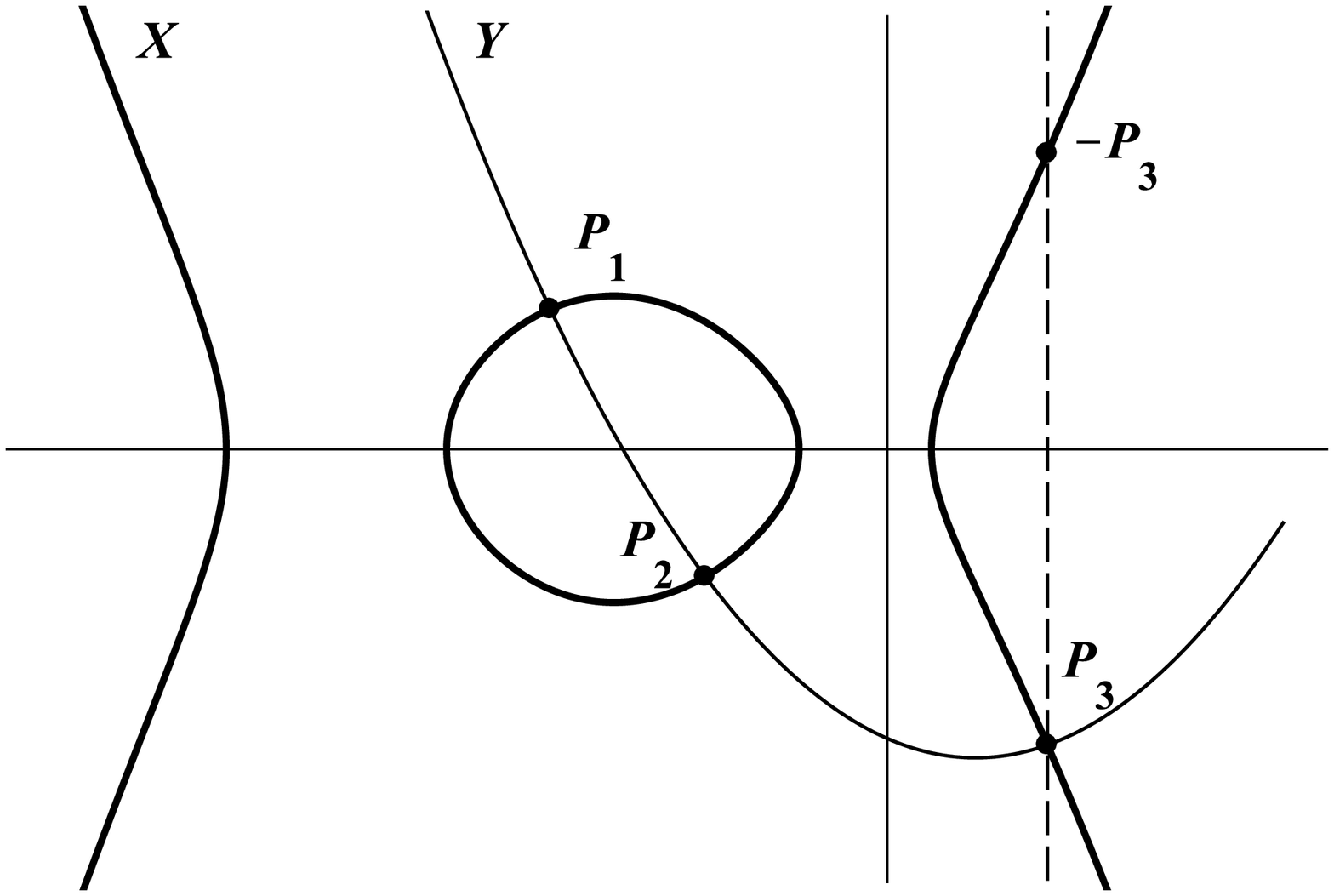} \\ a)  $(P_1+P_2)+P_3+P_\infty=0$}
\end{minipage}
\hfill
\begin{minipage}[h]{0.49\linewidth}
\center{\includegraphics[width=0.85\linewidth,height=0.2\textheight]{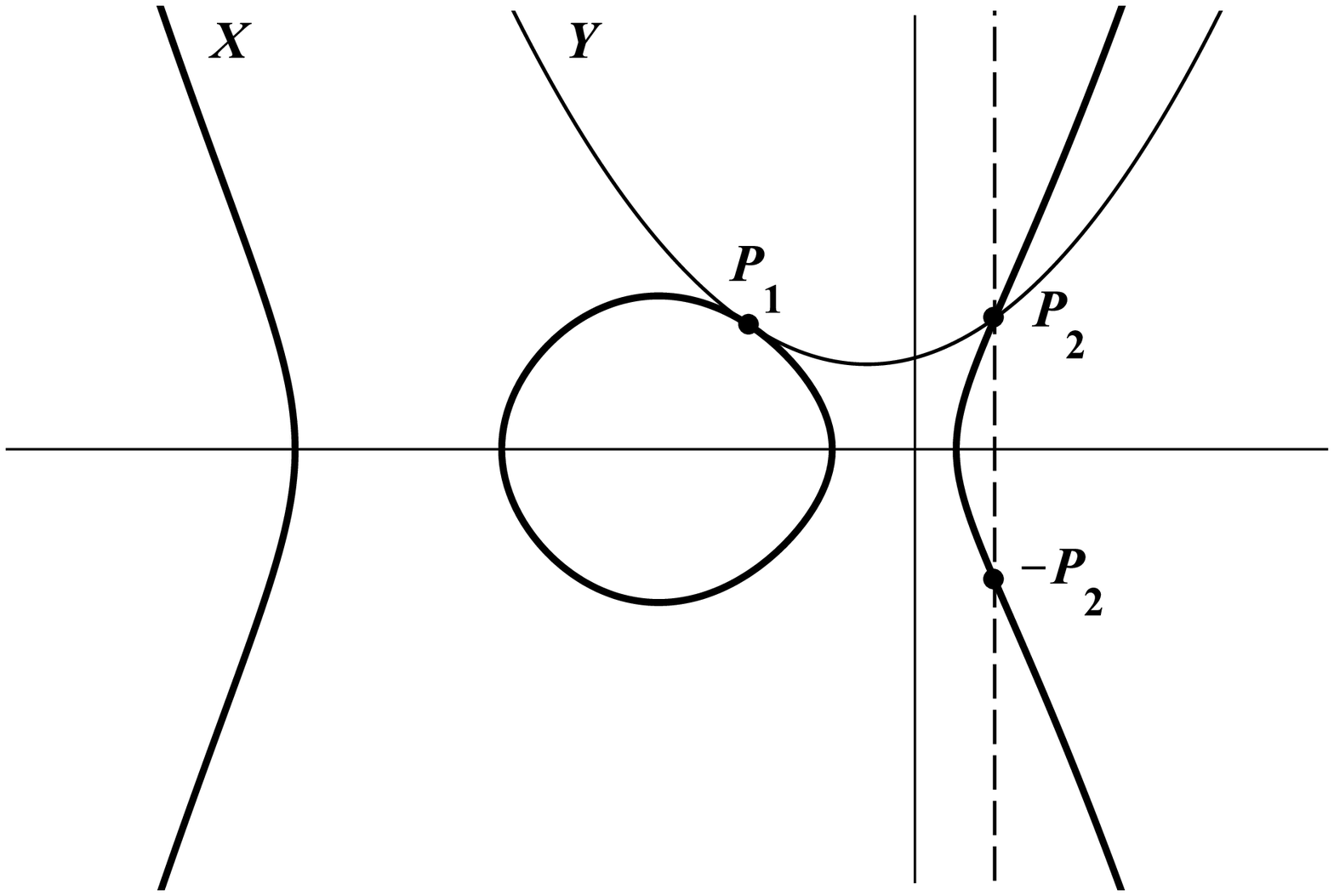} \\  b) $(2P_1)+P_2+P_\infty=0$}
\end{minipage}
\caption{Interection of  curve $X$  (\ref{ell-curve}) with parabola $Y:\,y=\sqrt{a_4}x^2+b_1x+b_0$}
\end{figure}

At $b_2\neq \sqrt{a_4}$ support of the intersection divisor consists of four  points $P_i\neq P_\infty$ up to multiplicity. Let us consider the following partitions of this divisor
\[
div(X\cdot Y)=(P_1+P_2+P_3)+P_4\,,\quad div(X\cdot Y)=(2P_1+P_2)+P_3\,,\quad
div(X\cdot Y)=(3P_1)+P_2\,,
\]
see Figure 2.

\begin{figure}[!ht]
\begin{minipage}[h]{0.32\linewidth}
\center{\includegraphics[width=0.85\linewidth, height=0.2\textheight]{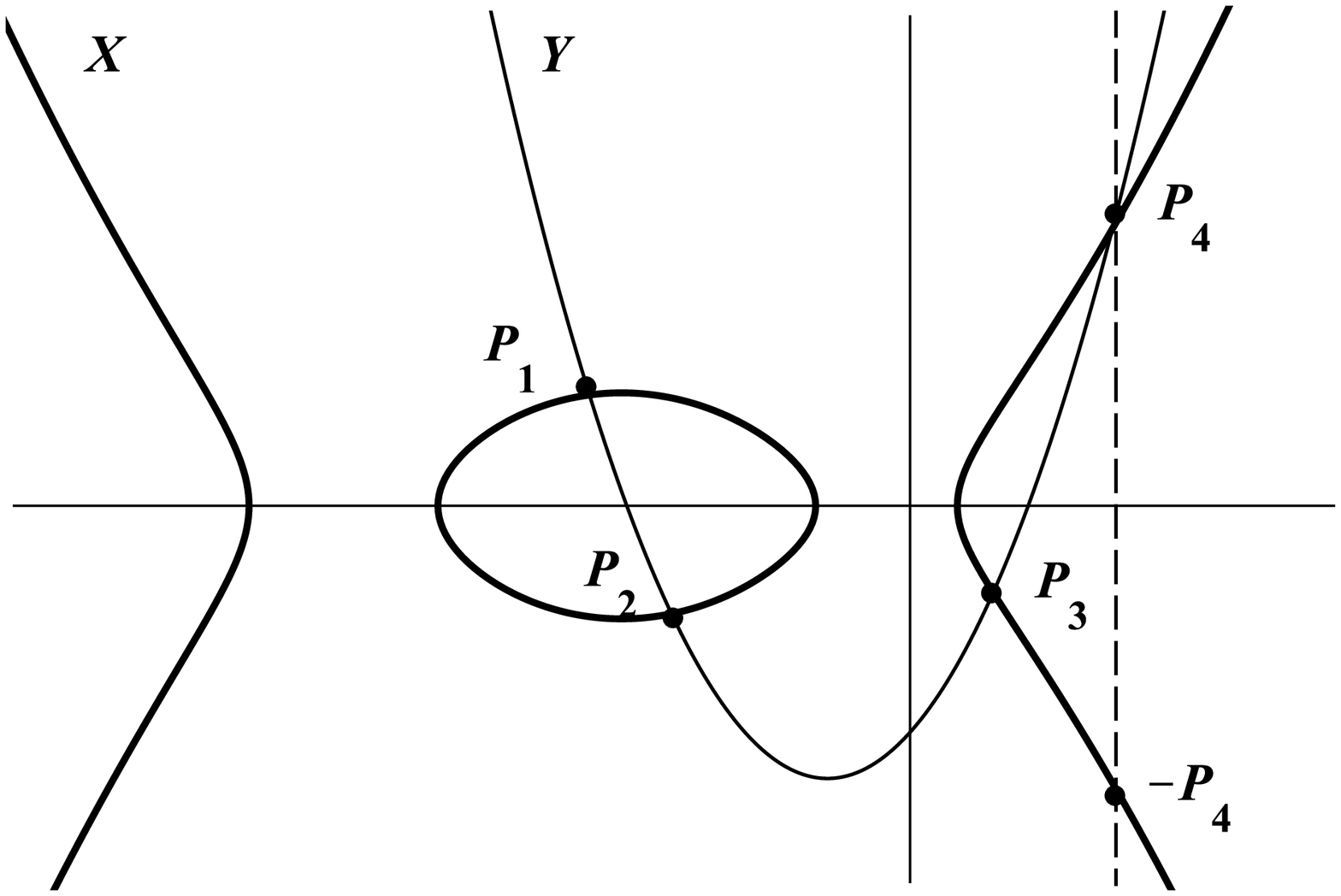} \\  a) $(P_1+P_2+P_3)+P_4=0$}
\end{minipage}
\hfill
\begin{minipage}[h]{0.32\linewidth}
\center{\includegraphics[width=0.85\linewidth,height=0.2\textheight]{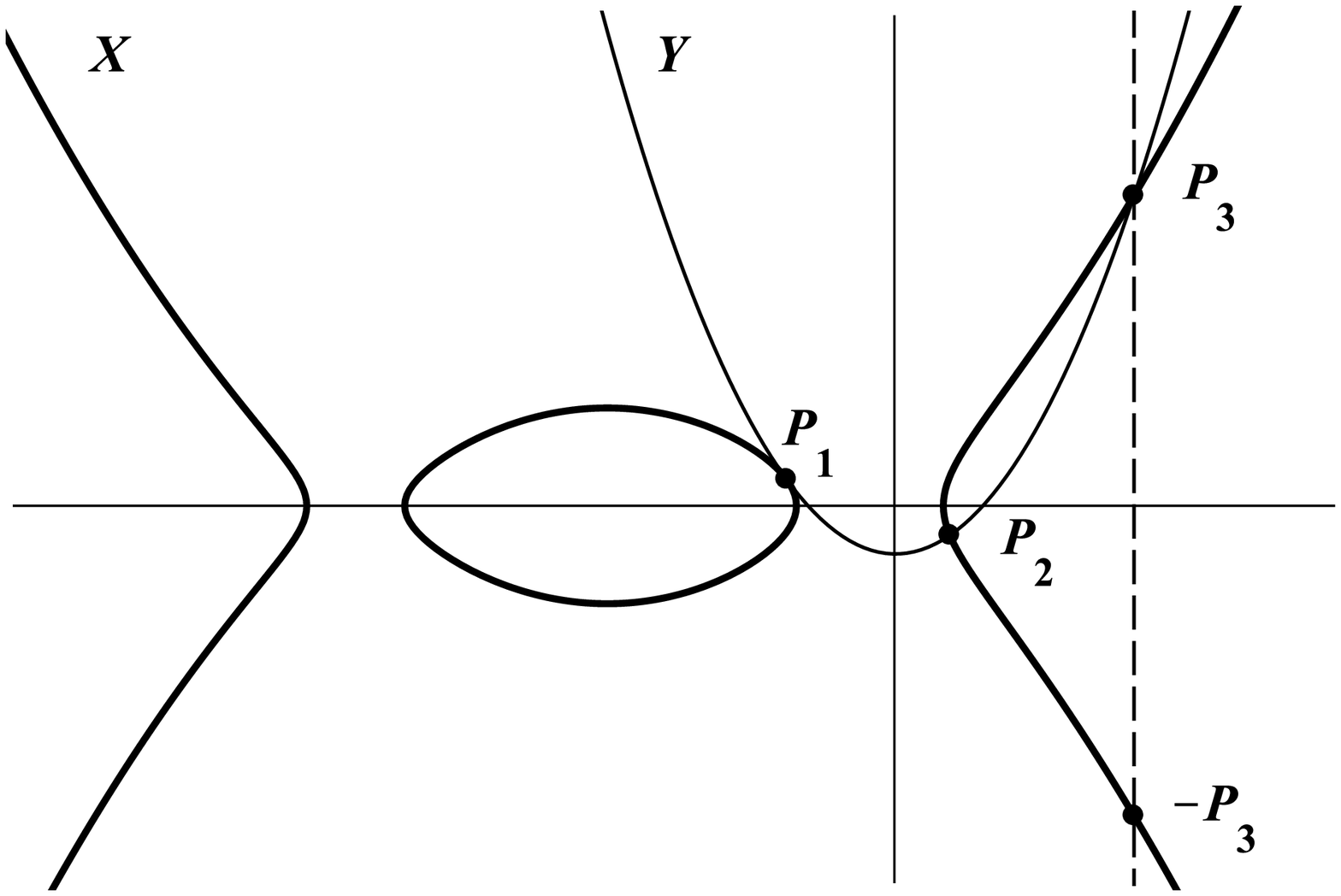} \\  b) $(2P_1+P_2)+P_3=0$}
\end{minipage}
\hfill
\begin{minipage}[h]{0.32\linewidth}
\center{\includegraphics[width=0.85\linewidth,height=0.2\textheight]{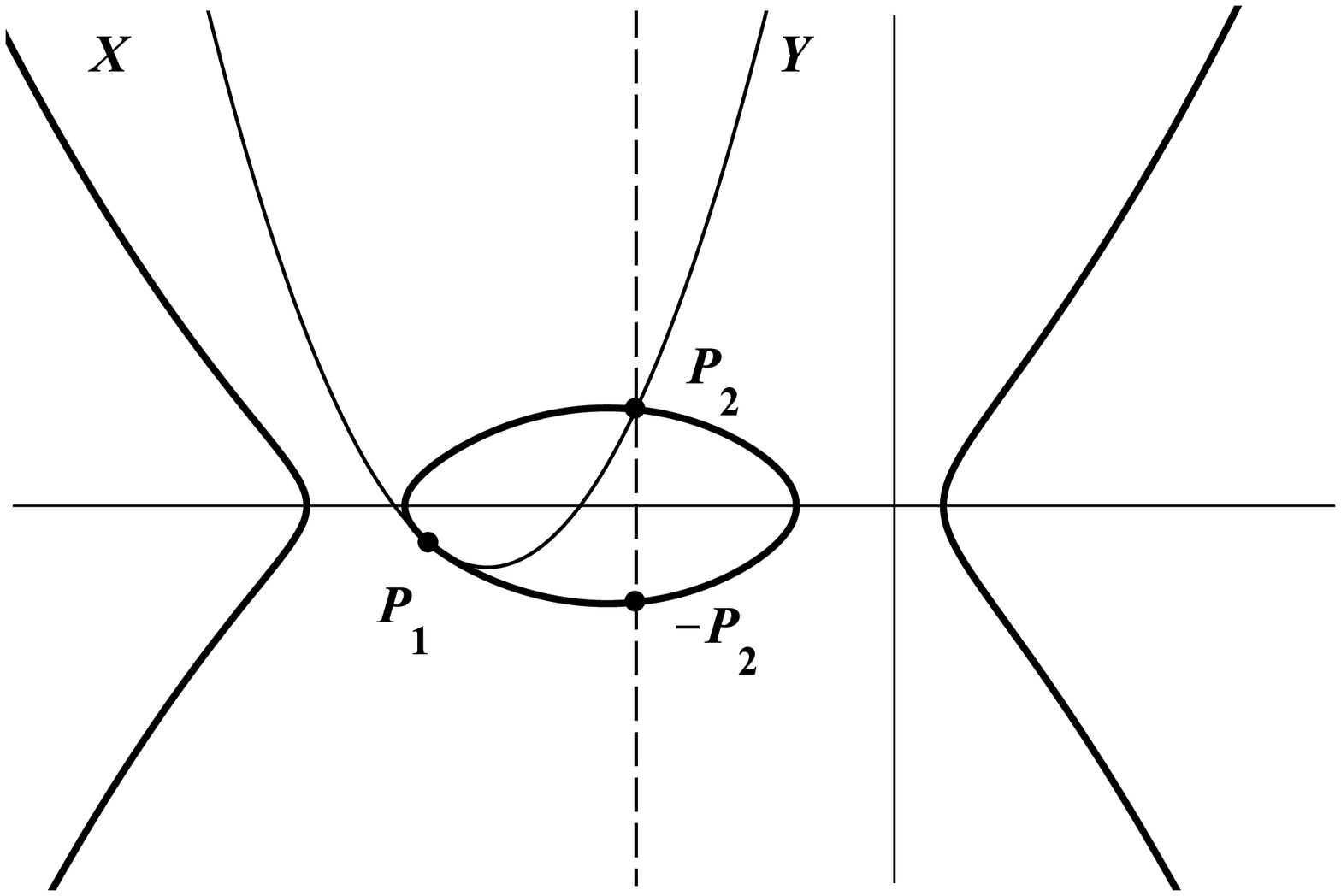} \\  c) $(3P_1)+P_2=0$}
\end{minipage}
\caption{Intersection  of  curve $X$  (\ref{ell-curve}) with parabola $Y:\,y=b_2x^2+b_1x+b_0$}
\end{figure}

In the first case parabola $Y$ is defined by the Lagrange interpolation  using three ordinary points $P_1, P_2$ and $P_3$. In the second and third cases  parabola $Y$ is defined by the Hermite interpolation using either double  and ordinary  points $2P_1, P_2$  or one triple point $3P_1$, respectively.

In the first case abscissa of the fourth intersection point is
\bq\label{add-ell4}
x_4=-x_1-x_2-x_3+\varphi(x_1,x_2,x_3,y_1,y_2,y_3)\,,\qquad\varphi=-\dfrac{a_3-2b_1b_2}{a_4-b_2^2}\,,
\eq
where function $\varphi$  is defined using coefficients of  quadratic polynomial  $\mathcal P(x)=b_2x^2+b_1x+b_0$
\bq\label{lag-cub}
\mathcal P(x)=\dfrac{(x-x_2)(x-x_3)y_1}{(x_1-x_2)(x_1-x_3)}
+\dfrac{(x-x_1)(x-x_3)y_2}{(x_2-x_1)(x_2-x_3)}+\dfrac{(x-x_1)(x-x_2)y_3}{(x_3-x_1)(x_3-x_2)}\,.
\eq

In the second case expression for the abscissa looks like
\[
x_3=-2x_1-x_2+\varphi(x_1,x_2,y_1,y_2)\,,\qquad \varphi=-\dfrac{a_3-2b_1b_2}{a_4-b_2^2}\,.
\]
Here function  $\varphi$ is defined via  coefficients of the same polynomial  $\mathcal P(x)=b_2x^2+b_1x+b_0$
and Hermite interpolation formulae
\[
\mathcal P(x)=\frac{(x-x_1)^2y_2-(x-2x_1+x_2)(x-x_2)y_1}{(x_1-x_2)^2}+\dfrac{(x-x_1)(x-x_2)(4a_4x_1^3+3a_3x_1^2+2a_2x_1+a_1)}{2y_1(x_1-x_2)}
\]

In the third case, when we consider tripling  the prime divisor on  $X$
\[
(x_2,y_2)=3(x_1,y_1)\,,
\]
second abscissa is equal to
\bq\label{trip-gen}
x_2= -3x_1+\varphi(x_1,y_1)\,,\qquad\varphi=-\dfrac{a_3-2b_1b_2}{a_4-b_2^2}\,,
\eq
where function $\varphi$ is defined  via  coefficients of the polynomial
\bq\label{trip-pol}
\begin{array}{rcl}
\mathcal P(x)=b_2x^2+b_1x+b_0&=&-\dfrac{(x-x_1)^2 (4 a_4 x_1^3+3 a_3 x_1^2+2 a_2 x_1+a_1)^2}{8 y_1^3}\\
\\
&+&\dfrac{(x-x_1)\Bigl(x \bigl(6 a_4 x_1^2+3 a_3 x_1+a_2\bigr)-2 a_4 x_1^3+a_2 x_1+a_1\Bigr)}{2 y_1}+y_1\,.
\end{array}
\eq

In the generic cases, using intersection divisors of plane curve $X$ with  auxiliary curves
\[
Y:\qquad y=b_Nx^N+b_{N-1}x^{N-1}+\cdots+b_0\,,\qquad N=1,2,3,\ldots
\]
we can describe multiplication of the prime divisor on  integer$P_1=nP_2$, which is a key ingredient of the modern elliptic curve cryptography, and other configurations of the prime divisors.

\subsection{Examples of finite-difference equations}
Let us identify curve $X$ with one of the common level curves of  $K$ (\ref{eul-k}) and $T$ (\ref{eul-h}) and
consider intersection divisor
\[div(X\cdot Y)=(P_1+P_2)+P_3+P_\infty\]
of this curve with parabola $Y$, see Figure 1a. Substituting
\[\begin{array}{rll}
&x_1=M_3\,,\qquad &y_1=(\a_2-\a_1)M_1M_2\,,\\ \\
&x_2=\widetilde{M}_3\,,\qquad &y_2=(\a_2-\a_1)\widetilde{M}_1\widetilde{M}_2\,,\\ \\
&x_3=\widehat{M}_3\,,\qquad &y_3=(\a_2-\a_1)\widehat{M}_1\widehat{M}_2\,,\\
\end{array}
\]
where  $M_j$, $\widetilde{M}_j$ and  $\widehat{M}_j$  are solutions of equations (\ref{eul-eq}) at $t=t_1,t_2,t_3$, in (\ref{add-gen})   one gets a 3-point finite-difference equation
\begin{eqnarray}\label{M3-add}
M_3+\widetilde{M}_3+\widehat{M}_3&=&\dfrac{\sqrt{a_4}\left(M_3^2-\widetilde{M}_3^2\right)
+(\a_1-\a_2)\left(M_1M_2-\widetilde{M}_1\widetilde{M}_2\right)}{2\sqrt{a_4}\left(M_3-\widetilde{M}_3\right)}\\
\nn\\
&+&\dfrac{(2a_4M_3\widetilde{M}_3-a_2)\left(M_3-\widetilde{M}_3\right)
+2\sqrt{a_4}(\a_1-\a_2)\left(M_1M_2\widetilde{M}_3-\widetilde{M}_1\widetilde{M}_2M_3\right)}{
2a_4\left(M_3^2-\widetilde{M}_3^2\right)+2\sqrt{a_4}(\a_1-\a_2)\left(M_1M_2-\widetilde{M}_1\widetilde{M}_2\right)}\,,\nn
\end{eqnarray}
whereas expression $y_3=-\mathcal P(x_3)$ gives rise to another finite-difference equation
\[\begin{array}{l}
(M_3-\widetilde{M}_3)(M_3-\widehat{M}_3)(\widetilde{M}_3-\widehat{M}_3)=\frac{(a_2-a_1)\left((\widetilde{M}_1 \widetilde{M}_2- \widehat{M}_1 \widehat{M}_2)M_3+(\widehat{M}_1 \widehat{M}_2-M_1 M_2) \widetilde{M}_3+(M_1 M_2- \widetilde{M}_1 \widetilde{M}_2)\widehat{M}_3 \right)}{\sqrt{a_4}}\,,
\end{array}
\]
which is compatible with (\ref{M3-add}). Here coefficients $a_4$ and $a_2$  of $X$ are given by  (\ref{eul-coeff}).  These equations  explicitly define   non-invertible multivalued map
\[
\left(
  \begin{array}{c}
    M_1,\,\widetilde{M_1} \\
    M_2,\,\widetilde{M_2} \\
    M_3 ,\,\widetilde{M_3}\\
  \end{array}
\right)\to
\left(
  \begin{array}{c}
    \widehat{M_1} \\
    \widehat{M_2} \\
    \widehat{M_3}\\
  \end{array}
\right)\,,
\]
which preserves the first integrals and the Poisson bracket. To calculate Poisson brackets between  $\widehat{M}_i$ we have to take into account that $h=2T$ has nontrivial Poisson brackets with ${M}_i$ and $\widetilde{M}_i$, see discussion in \cite{bob98,fed05}.

If we put
 \[\begin{array}{rll}
x_1&=M^{(k)}_3\,,\qquad &y_1=(\a_2-\a_1)M^{(k)}_1M^{(k)}_2\,,\\ \\
x_2&=\lambda_k\,,\qquad &y_2=\pm\sqrt{f(\lambda_k)}\,,\qquad \lambda_k\in\mathbb C\,,\\ \\
x_3&={M}^{(k+1)}_3\,,\qquad &y_3=(\a_2-\a_1)M^{(k+1)}_1M^{(k+1)}_2\,,\\
\end{array}
\]
where $f(x)$ is a function on the phase space defined by (\ref{eul-deq}-\ref{eul-coeff}), into the same expressions   (\ref{add-gen}) and $y_3=-\mathcal P(x_3)$,  one gets a recurrence chain of 2-point invertible maps
 \[
\cdots\,\xrightarrow[\lambda_{k-2}]{}\left(
  \begin{array}{c}
    M_1^{(k-1)} \\
    M_2^{(k-1)} \\
    M_3^{(k-1)}\\
  \end{array}\right)\xrightarrow[\lambda_{k-1}]{}\left(
  \begin{array}{c}
    M_1^{(k)} \\
    M_2^{(k)} \\
    M_3^{(k)}\\
  \end{array}
\right)
\xrightarrow[\lambda_{k}]{}\left(
  \begin{array}{c}
    M_1^{(k+1)} \\
    M_2^{(k+1)} \\
    M_3^{(k+1)} \\
  \end{array}
\right)\xrightarrow[\lambda_{k+1}]{}\,\cdots
\]
depending on parameters of discretization $\lambda_1,\lambda_2,\ldots$.  Similar to the Moser-Veselov correspondence (\ref{eul-mv}) and the Fedorov discretization (\ref{eul-fed}) these maps do not explicitly involve a time step $t_j-t_{j+1}$. Straightforward calculations allow us to prove that  these 2-point maps preserve first integrals $K,T$ (\ref{eul-k},\ref{eul-h}) and the Poisson  bracket (\ref{eul-pb}).

Doubling of prime divisor
\[-P_2=2P_1
\]
at
 \[
 x_1=M_3\,,\quad y_1=(\a_2-\a_1)M_1M_2\,,\qquad
 x_2=\widetilde{M}_3\,,\quad y_2=(\a_2-\a_1)\widetilde{M}_1\widetilde{M}_2
 \]
in  (\ref{doub-ell})  generates   map $M_i\to\widetilde{M}_i$:
\bq\label{M123-d}
\begin{array}{rcl}
\widetilde{M}_1&=&-\dfrac{M_1(\b_1M_2^2+\b_3M_3^2)}{2\sqrt{\b_1\b_3}M_2M_3}+\dfrac{\b_2M_2M_3}{2\sqrt{\b_1\b_3}M_1}\,,\\
\\
\widetilde{M}_2&=&\phantom{-}\dfrac{M_2(\b_1M_1^2+\b_2M_3^2)}{2\sqrt{\b_1\b_2}M_1M3}-\dfrac{\b_3M_1M_3}{2\sqrt{\b_1\b_2}M_2}\,,\\
\\
\widetilde{M}_3&=&-\dfrac{M_3(\b_3M_1^2+\b_2M_2^2)}{2\sqrt{\b_2\b_3}M_1M_2}+\dfrac{\b_1M_1M_2}{2\sqrt{\b_2\b_3}M_3}\,,\\
\end{array}
\eq
where
\[
\b_1= \a_1-\a_2\,,\qquad \b_2 = \a_2-\a_3\,,\qquad  \b_3 = \a_3-\a_1\,.
\]

\begin{prop}
Two point  map $M_i\to\widetilde{M}_i$ (\ref{M123-d}) preserves the form of first integrals $K,T$ (\ref{eul-k},\ref{eul-h}) and doubles the Poisson bracket (\ref{eul-pb}), i.e.
\[
\{M_i,M_j\}=\varepsilon_{ijk}M_k\,,\quad\Rightarrow\quad \{\widetilde{M}_i,\widetilde{M}_j\}=2\varepsilon_{ijk}\widetilde{M}_k\,.
\]
Map $M_j\to L_j=\widetilde M_j/2$  preserves the original Poisson bracket (\ref{eul-pb})
\[
\{M_i,M_j\}=\varepsilon_{ijk}M_k\,,\quad\Rightarrow\quad \{L_i,L_j\}=\varepsilon_{ijk}L_k
\]
and multiplies the  first integrals on the scaling factor due to e homogeneity of integrals of motion
\[\begin{array}{rcl}
H&=&\a_1M_1^2+\a_2M_2^2+\a_3M_3^2=4 \left(\a_1L_1^2+\a_2L_2^2+\a_3L_3^2\right)\,,\\
\\
K&=&M_1^2+M_2^2+M_3^2=4\left(L_1^2+L_2^2+L_3^2\right)\,.
\end{array}
\]
\end{prop}
The proof is a straightforward calculation.

Similarly we can obtain more complicated finite-difference equations using intersection divisors with four  points, see Figure 2. For instance, at
\[\begin{array}{rcl}
&x_1=M_3\,,\quad y_1=(\a_2-\a_1)M_1M_2\,,\qquad
&x_2=\widetilde{M}_3\,,\quad y_2=(\a_2-\a_1)\widetilde{M}_1\widetilde{M}_2\,,\\ \\
&x_3=\widehat{M}_3\,,\quad y_3=(\a_2-\a_1)\widehat{M}_1\widehat{M}_2\,,\qquad
&x_4=\widehat{\widetilde M}_3\,,\quad y_4=(\a_2-\a_1)\widehat{\widetilde M}_1\widehat{\widetilde M}_2\,,
\end{array}
\]
equations (\ref{add-ell4}), (\ref{lag-cub}) give rise to the following  finite-difference equation
\bq\label{eu-fd4}
M_3+\widetilde{M}_3+\widehat{M}_3+\widehat{\widetilde{M}}_3=\varphi(M,\widetilde{M},\widehat{M})\,,
\eq
where  function $\varphi(M,\widetilde{M},\widehat{M})=\varphi_1/\varphi_2$ is given by
\[\begin{array}{rcl}
\varphi_1&=&2(\a_1-\a_2)^2\Bigl(M_1M_2(\widehat{M}_3-\widetilde{M}_3)+\widetilde{M}_1 \widetilde{M}_2(M_3-\widehat{M}_3)+\widehat{M}_1 \widehat{M}_2 (\widetilde{M}_3-M_3)\Bigr)\\
\\
&\times&\Bigl(M_1M_2(\widehat{M}_3^2-\widetilde{M}_3^2)+\widetilde{M}_1 \widetilde{M}_2(M_3^2-\widehat{M}_3^2)+\widehat{M}_1 \widehat{M}_2 (\widetilde{M}_3^2-M_3^2)\Bigr)\,,
\end{array}
\]
and
\[
\begin{array}{rcl}
\varphi_2&=&(\a_1-\a_2)^2\Bigl(M_1M_2(\widehat{M}_3-\widetilde{M}_3)+\widetilde{M}_1 \widetilde{M}_2(M_3-\widehat{M}_3)+\widehat{M}_1 \widehat{M}_2 (\widetilde{M}_3-M_3)\Bigr)\\
\\
&+&(\a_2-\a_3)(\a_1-\a_3)(M_3-\widetilde{M}_3)^2(M_3-\widehat{M}_3)^2(\widetilde{M}_3-\widehat{M}_3)^2\,.
\end{array}
\]
Straightforward calculations allow us to check that  this 4-point non invertible map
\[
\left(
  \begin{array}{c}
    M_1,\,\widetilde{M_1},\,  \widehat{M_1} \\
    M_2,\,\widetilde{M_2},\, \widehat{M_2} \\
    M_3 ,\,\widetilde{M_3},\, \widehat{M_3}\\
  \end{array}
\right)\to
\left(
  \begin{array}{c}
    \widetilde{\widehat{M}_1} \\
   \widetilde{\widehat{M}_2} \\
    \widetilde{\widehat{M}_3}\\
  \end{array}
\right)
\]
preserves first integrals $K,T$ (\ref{eul-k},\ref{eul-h}) and the Poisson  bracket (\ref{eul-pb}).

 Substituting parameter $\lambda_k$ instead  of variable  $\widehat{M_3}$ into (\ref{eu-fd4})  one gets a 3-point map  depending on  parameter of discretization  $\lambda_k$
 \[
\left(
  \begin{array}{c}
    M_1,\,\widetilde{M_1} \\
    M_2,\,\widetilde{M_2} \\
    M_3 ,\,\widetilde{M_3}\\
  \end{array}
\right)\xrightarrow[\lambda_k]{}
\left(
  \begin{array}{c}
    \widehat{\widetilde M_1} \\
    \widehat{\widetilde M_2} \\
    \widehat{\widetilde M_3}\\
  \end{array}
\right)\,,
\]
  If we take
$x_2=\lambda_{1k}$ and  $x_3=\lambda_{2k}$ instead of  $x_2=\widetilde{M}_3$ and $x_3=\widehat{M}_3$,  we obtain a  recurrence chain of 2-point invertible maps
 \[
\left(
  \begin{array}{c}
    M_1 \\
    M_2 \\
    M_3\\
  \end{array}
\right)\xrightarrow[\lambda_{1k},\lambda_{2k}]{}
\left(
  \begin{array}{c}
    \widehat{\widetilde M_1} \\
    \widehat{\widetilde M_2} \\
    \widehat{\widetilde M_3}\\
  \end{array}
\right)\,,
\]
depending on  two parameters of discretization. These maps are algebraic multivalued  mappings  containing
the following functions on  phase space
\[
\mu_{ik}=\pm\sqrt{a_4\lambda_{ik}^4+a_2\lambda_{ik}^2+a_0}\,,\qquad \lambda_{ik}\in\mathbb C\,,
\]
where $a_4,a_2$ and $a_0$ are given by (\ref{eul-coeff}).

Let us come back to construction of single-valued  maps and substitute
  \[x_1=M_3\,,\quad y_1=(\a_2-\a_1)M_1M_2\,,\qquad
 x_2=\widetilde{M}_3\,,\quad y_2=(\a_2-\a_1)\widetilde{M}_1\widetilde{M}_2
 \]
 into the expressions (\ref{trip-gen}) and (\ref{trip-pol}) for tripling  a point on elliptic curve
\[-P_2=3P_1\,,
\]
see Figure 2c. As a result one gets the following  map $M_i\to\widetilde{M}_i$:
\bq\label{eul-trip}
\begin{array}{rcl}
\widetilde{M}_1&=&\phantom{-}3M_1-\dfrac{4M_1^3}{D}\Bigl(M_1^2\bigl(\b_1M_2^2-\b_3M_3^2\bigr)^2
-\b_2M_2^2M_3^2(\b_1M_2^2+\b_3M_3^2)\Bigr)\,,
\\
\\
\widetilde{M}_2&=&-3M_2+\dfrac{4M_2^3}{D}\Bigl(M_2^2\bigl(\b_1M_1^2-\b_2M_3^2\bigr)^2-\b_3M_1^2M_3^2(\b_1M_1^2+\b_2M_3^2)\Bigr)\,,
\\
\\
\widetilde{M}_3&=&-3M_3+\dfrac{4M_3^3}{D}\Bigl(M_3^2\bigl(\b_2M_2^2-\b_3M_1^2\bigr)^2-\b_1 M_1^2M_2^2(\b_3M_1^2+\b_2M_2^2)\Bigr)\,,
\end{array}
\eq
where
\[
D=M_3^4\bigl(\b_3M_1^2-\b_2M_2^2\bigr)^2-2\b_1M_1^2M_2^2M_3^2(\b_3M_1^2+\b_2M_2^2)+\b_1^2M_1^4M_2^4\,.
\]
\begin{prop}
Two point  map $M_i\to\widetilde{M}_i$  (\ref{eul-trip}) preserves first integrals $K,T$ (\ref{eul-k},\ref{eul-h}) and changes the Poisson bracket (\ref{eul-pb}) by the following rule
\[
\{M_i,M_j\}=\varepsilon_{ijk}M_k\,,\quad\Rightarrow\quad \{\widetilde{M}_i,\widetilde{M}_j\}=3\varepsilon_{ijk}\widetilde{M}_k\,.
\]
Map $M_j\to L_j=\widetilde M_j/3$  preserves the original Poisson bracket (\ref{eul-pb})
\[
\{M_i,M_j\}=\varepsilon_{ijk}M_k\,,\quad\Rightarrow\quad \{L_i,L_j\}=\varepsilon_{ijk}L_k
\]
and the first integrals up to the integer scaling factor
\[\begin{array}{rcl}
H&=&\a_1M_1^2+\a_2M_2^2+\a_3M_3^2=9 \left(\a_1L_1^2+\a_2L_2^2+\a_3L_3^2\right)\,,\\
\\
K&=&M_1^2+M_2^2+M_3^2=9\left(L_1^2+L_2^2+L_3^2\right)\,.
\end{array}
\]
\end{prop}
The proof is a straightforward calculation.

So, equations  (\ref{M123-d}) and (\ref{eul-trip}) give rise to a recurrence chain of 2-point mappings
\[
\cdots\,\xrightarrow[N]{}\left(
  \begin{array}{c}
    M_1^{(k-1)} \\
    M_2^{(k-1)} \\
    M_3^{(k-1)}\\
  \end{array}\right)\xrightarrow[N]{}\left(
  \begin{array}{c}
    M_1^{(k)} \\
    M_2^{(k)} \\
    M_3^{(k)}\\
  \end{array}
\right)
\xrightarrow[N]{}\left(
  \begin{array}{c}
    M_1^{(k+1)} \\
    M_2^{(k+1)} \\
    M_3^{(k+1})
  \end{array}
\right)\xrightarrow[N]{}\,\cdots
\]
with $N=2,3$, which can be considered as a counterpart of geometric progression. In contrast with 2-point algebraic maps from the Theorem 3.3 in \cite{fed05} these maps change original Poisson bracket and, therefore, they are different from the known 2-point maps preserving original bracket.

Because points on  plane curve $X$ (\ref{ell-curve}) are solutions of  equation (\ref{eul-deq}) at fixed time and, therefore, $M_i^{(k)}=M_i(t_k)$ are proportional to the Jacobi elliptic functions $sn(u_k)$, $cn(u_k)$, $dn(u_k)$, with $u_k = \mbox{const}\cdot t_k$, this recurrence chain represents a set of additional theorems for elliptic functions.

\section{Conclusion}
In \cite{bob98,fed05, mos91,sur03,ves88}  various Lax matrices for the Euler top, their refactorizations and the underlying addition theorems for the elliptic functions were used to obtain integrable 2-point finite-difference equations associated with motion of the  Euler top. Nowadays, refactorization in Poisson-Lie groups is viewed as one of the most universal mechanisms of integrability for integrable $2$-point maps. In this note, we come back to the  Abel construction of  addition theorems  in order to study $n$-point finite-difference equations sharing  integrals of motion and  Poisson bracket up to the integer scaling factor.

If a common level surface $X$ of integrals of motion can be  represented as a symmetric product of an algebraic curve (for instance, the spectral curve of the Lax matrix), we can apply both of these constructions to:
\begin{itemize}
  \item construction of integrable discrete maps \cite{hiet16,kuz02,mos91,sur03, ves88,ves91};
  \item study of relations between different integrable systems \cite{ts15a,ts15b,ts15c};
  \item construction of new integrable systems \cite{ts17a,ts17b,ts17c,ts18d};
  \item integration by quadratures.
\end{itemize}
Let us explain the last item by an example of the Steklov-Lyapunov system. The corresponding Kirchhoff equations of motion  were  solved explicitly by K\"{o}tter \cite{k00} using separation of variables and reduction of equations of motion to quadratures which have the form of the Abel-Jacobi map associated with a genus two hyperelliptic curve.  The main problem is that original phase variables are expressed in terms of separating variables in a complicated way and, therefore, we cannot use K\"{o}tter's results for analysis of motion in practice, see discussion in \cite{bm05,fed11,ts04}.

In \cite{ts12} we introduce other variables of separation associated with a simple  polynomial bi-Ha\-mil\-to\-ni\-an structure for  the Steklov-Lyapunov system. In this case original phase variables are easily expressed via variables of separation, which allows us to integrate by quadratures the nontrivial integrable generalization of the Steklov–Lyapunov system discovered by Rubanovsky. Using modern computer algebra systems we can prove that canonical transformation relating K\"{o}tter's variables and "new"\, variables of separation is the doubling divisor operation on genus two hyperelliptic curves discussed in \cite{ts18d}. Consequently, sometimes  we can apply divisor arithmetic to simplification of  integration by quadratures.

Another reason to conduct these calculations is related to consideration of finite-difference equations  (\ref{fd-eq}) relating points on the common level surface $X$ of first integrals, which can not be realized as a product of the plane algebraic curves. In this generic case when we do not know variables of separation or the  Lax matrices,  we can continue to study various configurations of points on algebraic surface $X$ in the framework of the standard intersection theory \cite{bak97,eh16,ful84, grif04,kl05}.

We are very grateful to the referees for  thorough analysis of the manuscript, constructive  suggestions and  proposed  corrections, which certainly lead to a more profound discussion of the results. The work was supported by the Russian Science Foundation (project  15-12-20035).

\end{document}